\documentclass[preprint,12pt]{elsarticle}
\usepackage{amssymb}
\usepackage{amsthm}
\usepackage{amsmath}
\usepackage{graphicx}
\usepackage{tabularx}
\usepackage{makecell}
\usepackage{xcolor}
\usepackage{multirow}
\usepackage{multicol}
\usepackage{booktabs}
\usepackage{diagbox}
\newcommand{\cmt}[1]{\textcolor{blue}{#1}}

\journal{Chaos, Solitons \& Fractals}
\begin{document}
\begin{frontmatter}
\title{Predicting the critical behavior of complex dynamic systems via learning the governing mechanisms}  
\author[a]{Xiangrong Wang\corref{cor1}}
\author[b]{Dan Lu}
\author[a]{Zongze Wu\corref{cor1}}
\author[c]{Weina Xu}
\author[d]{Hongru Hou}
\author[d]{Yanqing Hu}
\author[e,f]{Yamir Moreno\corref{cor1}}
\cortext[cor1]{Corresponding authors}
\address[a]{College of Mechatronics and Control Engineering, Shenzhen University,Shenzhen,518060,China}
\address[b]{School of Reliability and Systems Engineering, Beihang University,Beijing, China}
\address[c]{School of Computer Science and Engineering, Beihang University, Beijing, China}
\address[d]{Southern University of Science and Technology, Shenzhen, China}
\address[e]{Institute for Biocomputation and Physics of Complex Systems (BIFI), University of Zaragoza, Zaragoza, 50018, Spain}
\address[f]{Department of Theoretical Physics, University of Zaragoza, Zaragoza, 50009, Spain}
\begin{abstract}
Critical points separate distinct dynamical regimes of complex systems, often delimiting functional or macroscopic phases in which the system operates. However, the long-term prediction of critical regimes and behaviors is challenging given the narrow set of parameters from which they emerge. Here, we propose a framework to learn the rules that govern the dynamic processes of a system. The learned governing rules further refine and guide the representative learning of neural networks from a series of dynamic graphs. This combination enables knowledge-based prediction for the critical behaviors of dynamical networked systems. We evaluate the performance of our framework in predicting two typical critical behaviors in spreading dynamics on various synthetic and real-world networks. Our results show that governing rules can be learned effectively and significantly improve prediction accuracy. Our framework demonstrates a scenario for facilitating the representability of deep neural networks through learning the underlying mechanism, which aims to steer applications for predicting complex behavior that learnable physical rules can drive.
\end{abstract}

\end{frontmatter}

\section{Introduction}

The dynamics of contagion is a complex phenomenon whose outcome depends on several factors, ranging from the very nature of the contagion (e.g., social versus biological) to the means through which the spreading takes place. For instance, the dissemination of information is impacted by the design of the information-sharing platforms, the algorithms used to transmit, filter, or display such information and ultimately, by human behavior. Studies carried out in the last few years have allowed for insights into problems such as the outburst of information \cite{davis_phase_2020} and fake news~\cite{lazer2018science, apuke2021fake}, the widespread propagation of obesity and behaviors in human populations \cite{huang2016social,chen2021psychological} or the rise of social movements \cite{gonzalez-bailon_dynamics_2011}. On the other hand, the mathematical characterization of contagion processes among humans is also important because it is the framework used to study the spreading of diseases, which occurs through social interactions that are encoded by networks. Models such as the Susceptible-Infected-Recovered (SIR) model and its variants have been widely applied to study such spreading dynamics on  networks~\cite{kermack1927contribution,vespignani2012modelling,machado2022effect,DEARRUDA20181}. 

The variety of determinants of contagion processes could lead to dynamic outcomes that are associated with subcritical, critical, and supercritical regimes. A large body of the recent literature has focused on characterizing the conditions that drive a system in any of the previous regimes (see, for instance, \cite{DEARRUDA20181} and references therein). In the context of contagion processes, critical behaviors refer to the system-level transition from a local spreading state to a global spreading state that spans the entire network. These critical behaviors are commonly observed in both artificial and real-world networks, leading to the emergence of large-scale disease outbreaks and system-wide information cascades. Additionally, near the transition point between local and global spreading, the outcome of the dynamics remains elusive, even under identical system settings ~\cite{,hu2018local,dehning2020inferring,oliveira2021mathematical}. 

Thus, the accurate prediction of critical behaviors is of utmost theoretical and practical importance. Yet, it is a challenging task because they might appear for a narrow range of system parameters~\cite{pastor2015epidemic,gnanvi2021reliability}. To predict system-level behaviors, a variety of deep learning models have been proposed based on the representation learning of dynamical features from observed dynamic graphs. Deep learning models, especially graph neural networks~\cite{gori2005new,kipf2016semi} and recurrent neural networks~\cite{elman1990finding}, have been widely applied for traffic forecasting, social recommendation, anomaly detection, and air quality predictions, among others. However, the direct application of these models to predict critical behaviors could lead to inaccurate predictions, as these models tend to rely heavily on the short-range dependence of spatial and temporal information in the training data, which might suppress the representability of information between different system states. The difficulty in making predictions is further exacerbated because critical behaviors typically constitute a small proportion of observation data, with the possibility of discontinuous transitions with no precursors at all. Nonetheless, the main hypothesis is that given that governing rules provide temporal and spatially invariant knowledge on how the (micro)states of a system evolve, an effective representation of such governing rules could guide and refine the prediction of critical behaviors in dynamical systems. 

In this work, we introduce a graph neural network-based framework to effectively learn dynamical governing rules, through which to achieve an accurate prediction for critical behaviors. Our approach employs attentive graph neural networks to discern the underlying governing rules shaping the system's evolution. Leveraging the learned governing rules, we refine and guide a graph recurrent unit to capture long-term state dependencies. Using spreading dynamics on various network structures, we show that the framework accurately predicts two properties that characterize critical regimes widely found in complex social systems. Lastly, our framework demonstrates a scenario for enhancing the representability of deep neural networks through learning the underlying mechanism, which can be generalized to predict, for example, pattern formation~\cite{colizza2007reaction} and system instability~\cite{buldyrev2010catastrophic}.

\section{Related work}
Our framework is built upon graph neural networks and is designed to: 1) learn the governing rules from dynamic graphs and 2) learn effective representation for long-term dependencies. Below, we discuss related work to our method from these two perspectives.

\subsection{Deep learning from time series data}

One fundamental design of neural networks for time series forecasting is to effectively learn spatial and temporal knowledge from the training data. Recurrent neural networks \cite{elman1990finding} and their variants, such as the Long Short Term Memory (LSTM) \cite{hochreiter1997long}, the Gated Recurrent Unit (GRU)~\cite{cho2014properties}, make forecasting based on memory information learned from prior inputs. To reduce the complexity of traditional LSTM and prediction errors, ConvLSTM \cite{shi2015convolutional} introduced a convolution structure to better extract spatiotemporal knowledge. However, when forecasting under high uncertainty and in the long-term, these models show limited performance due to difficulties in propagating learned knowledge to farther nodes. To solve the long-term forecasting problem, transformer-based approaches have been proposed. Autoformer~\cite{wu2021autoformer} developed an encoder-decoder model with decomposition capabilities, and made attention approximations based on Fourier transform, while Informer~\cite{zhou2021informer} took a stacked multistage approach to encode long sequences. Moreover, to make predictions based on multivariate time series, LSTNet~\cite{lai2018modeling} introduced a recurrent-skip design to get long-term dependencies and combined Convolution Neural Networks and Recurrent Neural Networks to tune long-term and short-term patterns. Finally, CF-RNN~\cite{stankeviciute2021conformal} developed a conformal prediction algorithm for time series forecasting with high stakes.

\subsection{Deep learning for dynamic complex systems}

To process data that can be encoded in graphs, such as social networks and infrastructure networks, Graph Convolutional Networks (GCN)~\cite{kipf2016semi} and Graph Attention Networks (GAT)~\cite{velickovic2017graph} are typical models to learn the hidden spatial knowledge of nodes. Based on GCN, STGCN~\cite{guo2019attention} further introduced spatial and temporal attention mechanisms to capture the dynamic spatial and temporal correlations in traffic data. Additionally, to improve the performance and robustness of traffic flow prediction while abandoning pre-defined graphs, AGCRN~\cite{bai2020adaptive} aggregated GCN, Data Adaptive Graph Generation (DAGG) module, and GRU to capture node-specific spatial and temporal correlations in time-series data. Finally, StemGNN~\cite{cao2020spectral} combined Graph Fourier Transform and Discrete Fourier Transform to better model the intra-series temporal patterns and inter-series correlations jointly, whereas DSTAGNN~\cite{lan2022dstagnn} proposed a new dynamic spatiotemporal aware graph based on a data-driven strategy.

More related to the present contribution, it is also worth mentioning that with the emergence of the COVID-19 pandemic, epidemic forecasting methods based on deep neural networks were extensively developed. For instance, MPNN-LSTM~\cite{panagopoulos2021transfer} exploited an LSTM architecture to integrate the district features obtained from graph convolutional layers, enabling the transfer of learned knowledge to different countries for global forecasting. On the other hand, REGENN~\cite{spadon2021pay} devised a graph-inspired learning-representation layer and a neural network architecture for modeling spatial and temporal dynamic processes over different phases and CausalGNN \cite{wang2022causalgnn} introduced a casual module to learn additional epidemiological information from differential equations.

\section{Methods}
Here, we propose a graph neural network-based framework to predict critical behaviors of complex systems based on a series of dynamic graphs. Taking the spreading dynamics as an example, we predict the system's critical properties based on effective learning of microscopic governing rules of the spreading dynamics, i.e., the rules underpinning the spreading from an infected individual to susceptible individuals via its local network connections, and the effective learning of the macroscopic long-range dependence in dynamic graphs.

\subsection{Overview}
\begin{figure}[!t]
    \centering
    \includegraphics[width=\linewidth]{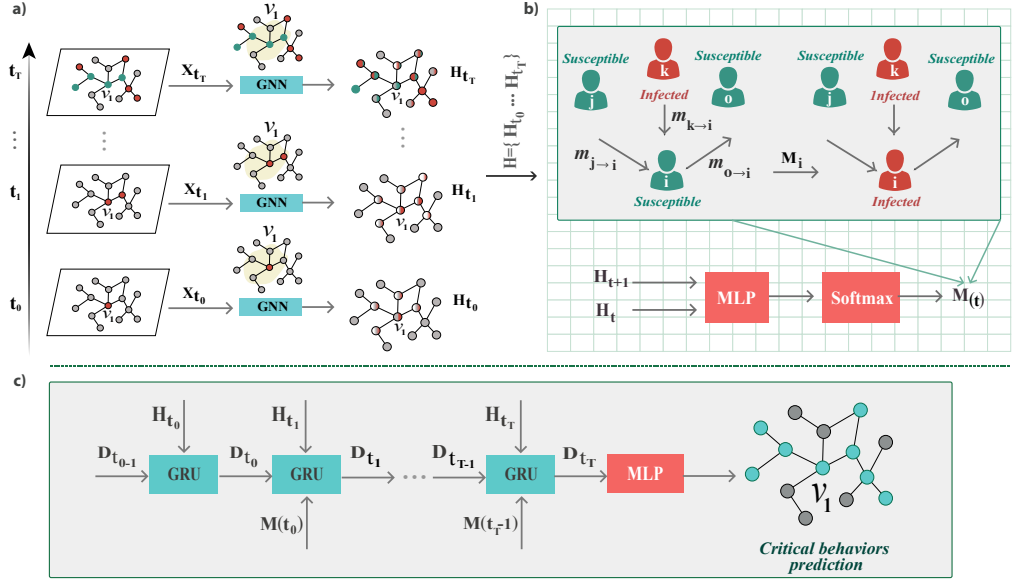}
    \caption{Architecture of graph neural network based framework to learn governing rules for the prediction of critical behaviors, which consists of {\it i)} a graph neural network based module (a) to learn the governing rules of spreading dynamics (b) and {\it ii)} a gated recurrent module (c) to learn long-range dependence.}
    \label{fig:architecture}
\end{figure}

\vspace{0.2cm}
The main framework, as shown in Figure~\ref{fig:architecture}, consists of two main modules: {\it i)} graph neural networks-based module with the designed attention mechanism to learn the governing rules of dynamic transitions and {\it ii)} the grated recurrent unit module to learn the long-range system state correlations from dynamic graphs. The combination of the two modules provides mechanism-based predictions for forecasting critical behaviors.

The first module deals with the rules governing the transition from one dynamic graph to the other. Let $G_t=(\mathcal{V},\mathcal{E},X_t)$ denote a dynamic graph, where $\mathcal{V}=\{v_1,\dots,v_N\}$ is the set of nodes, and $\mathcal{E}\subseteq |\mathcal{V}| \times |\mathcal{V}|$ is the set of edges connecting nodes. $X_t \in \mathbb{R}^{N\times 1}$ is a vector encoding nodal states at time $t$, which is designed to take discrete values of ${0, 1, -1}$ representing nodal states of susceptible, infected or recovered, respectively (see next section for more details). Specifically, the state of a node $v_i$ at time $t$ is represented by $X_{v_i,t} $. Each dynamic graph contains individuals' temporal states and their local network connections which channel the spreading of information.

The states of the nodes dynamically evolve according to the microscopic governing rules hidden in a series of dynamic graphs. Explicit extraction of decisive governing rules from dynamic graphs is necessary for the accurate prediction of the system's critical behavior. To this end, our model first learns spatial and temporal knowledge from dynamic graphs via a GNN-based module and then leverages this learned knowledge to guide and refine the prediction of the next system state. The graph features $X_t \in \mathbb{R}^{N\times 1}$ at time step $ t $ are encoded in $H_t \in \mathbf{R}^{N\times d}$ via graph neural networks, which integrate the states of neighboring nodes to extract spatial knowledge. In particular, we formulate an attention mechanism to highlight the most dynamically relevant neighbors from all local network connections. The pattern of governing rules to guide dynamic evolution can be recorded in adjacent timestamps. Therefore, the governing rules are designed to learn from combining the adjacent output $H_t$ and $H_{t+1}$. 

\begin{figure}[!t]
    \centering
   \includegraphics[width=\linewidth]{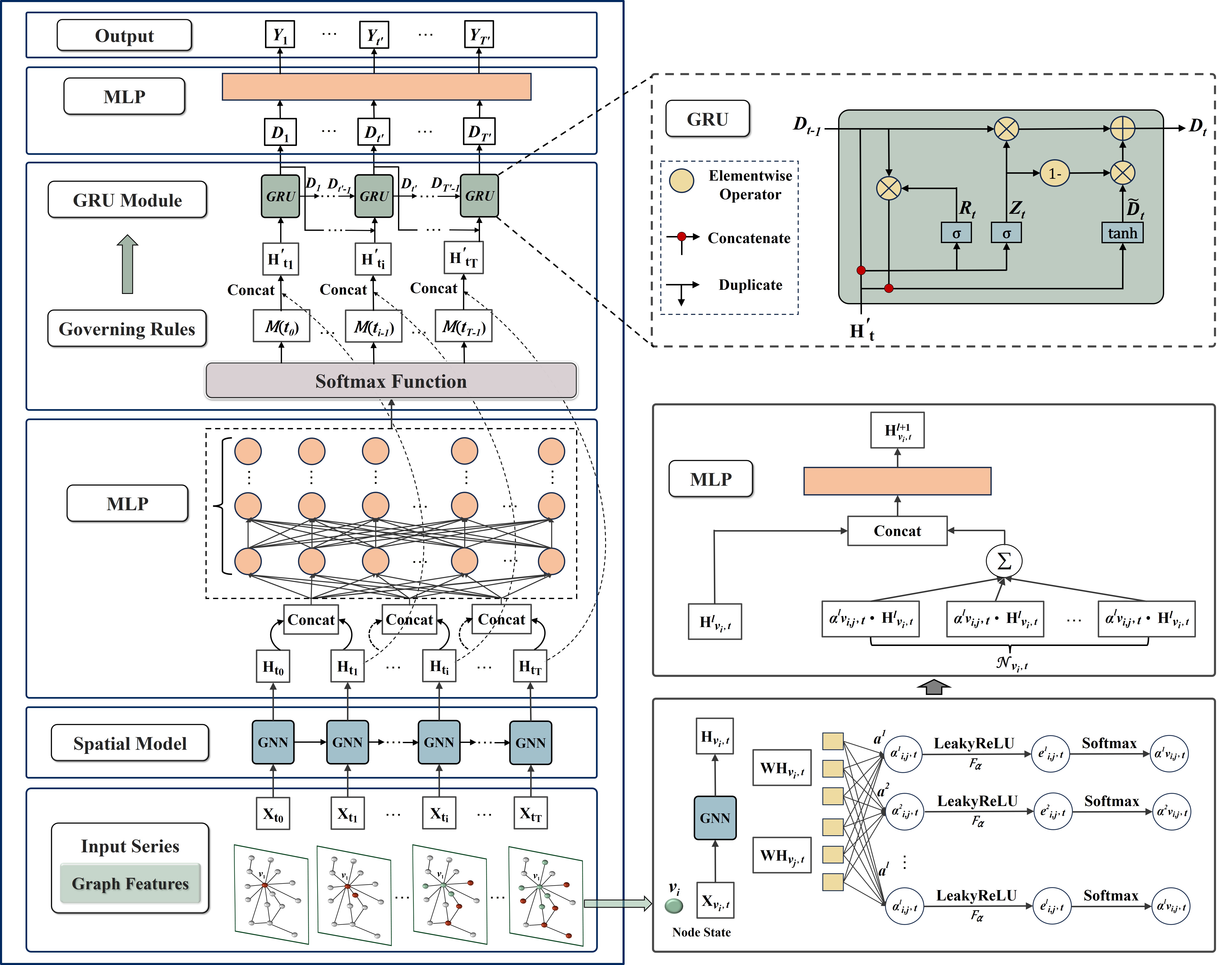}
    \caption{Prediction Framework}
    \label{fig:framework}
\end{figure}

The second module is designed to learn long-range system dependencies. We formulate a collection of integrated features of dynamic graphs, $H=\{H_{t_0},\dots, H_{t_i}, \ldots, H_{T}\}$ where $H_{t_i}$ extracts spatial knowledge at time $t_i$, into the Gated Recurrent Unit (GRU) module with learned governing rules that can directly guide the direction of evolution. The aim is to extract the temporal-dependent knowledge and distill the evolutionary trend hidden in dynamic graphs. Finally, critical behaviors $\hat{Y}$ are predicted by the output of the GRU model that covers spatial and temporal knowledge as shown in Figure~\ref{fig:framework}. Next, we show that our model can effectively learn the microscopic governing rules and thereby accurately predict critical behaviors.

\subsection{Learn governing rules of spreading dynamics}
\label{LFND}
\vspace{0.2cm}
In basic contagion processes, individuals who are initially infectious (or hold a piece of news for information spreading) can infect susceptible neighbors in their local neighborhood.  A cascade-like spreading process emerges and continues until no more infectious individuals are left. The final number of infected individuals (or people who know the news) determines the size of the outbreak. This simple dynamics is characterized by a critical transition between a regime in which there is no outbreak and a phase in which the outbreak involves a macroscopic fraction of the population.  The effective learning of the underlying governing rules, denoted as $\mathcal{M}(t)$, on how individuals spread the disease/information to others is therefore of paramount importance to accurately predict the long-term system's behavior.

Complex systems are characterized by the fact that the state of an individual node can dynamically evolve in response to its intrinsic behavior (such as self-recovery, self disinterest) and due to the peer influence received from its local connections (e.g., social pressure). The Susceptible-Infected-Recovered (SIR) spreading model is widely used to describe the above-described dynamic process, in which a susceptible individual can be infected by infectious neighbors with a probability $\beta \in [0,1]$ per contact. Moreover, in SIR disease dynamics, an infected individual recovers (or is removed) with a certain probability $\gamma \in [0,1]$, regardless of the interaction with its neighbors. Therefore, for one individual and depending on their current state, the probability of transitioning to another state at the next time step $t+1$ can be formulated as the probability of remaining in the original state and the probability of transferring to a new state, the sum of which is $1$. The probability for an individual node $v_i$  to remain in the same state at time $t+1$ can be written as

\begin{equation}\label{eq:governing_rules}
{\textrm{Pr}\left(X_{v_i,t+1} = X_{v_i,t}\right)} = \begin{cases}
(1- \beta )^{N_{v_{i,t}}^l}, &{\text{if}}\ {X_{v_i,t}=0} \\ 
{1-\gamma} &{{\text{if}}\ {X_{v_i,t}=1}} \\
{1} &{\text{otherwise.}} 
\end{cases}
\end{equation}

where $\beta$ is the probability for a susceptible node to be infected by a single contact with an infectious neighbor and $\gamma$ is the self-recovering probability for an infected node. $N_{v_{i,t}}^l$ refers to the number of infected neighbors of node $v_i$ at time $t$, which is a critical spatial and temporal feature to be learned. In addition, the probability for the node $v_i$ to transfer to a new state is obtained by $\textrm{Pr}\left(X_{v_i,t+1} = X_{v_i,t}\right)$.

The microscopic governing rule $\mathcal{M}(t)$ described in Equation \ref{eq:governing_rules} determines the spatial and temporal evolution of the system. We learn these mechanisms by capturing the spatial and temporal contacts of infected individuals from dynamic graphs using the GNN module. GNN enables the aggregation of neighboring features into the features $X_{v_{i},t}$ of an individual node $v_i$. The corresponding output is denoted as $H_{v_{i},t}$, which contains information about the spatial interactions between one individual and its neighboring nodes.

To effectively capture \emph{both} spatial and temporal knowledge for dynamic processes, we propose an aggregation function designed to preserve the intrinsic self-feature while aggregating neighboring features. In addition, since infected neighbors have a more direct impact on governing rules compared to all neighbors in local connections, as illustrated in Equation \ref{eq:governing_rules}, we leverage an attention mechanism in graph neural networks to assign higher weights to infected neighbors. The feature-update process for the node $v_{i,t}$ at the $l$-th iteration layer is formulated as:
\begin{equation}
\label{GAT}
\begin{aligned}
\alpha_{v_{i,j},t}^{l} &=\frac{{\rm exp}( \mathcal{F_\alpha} ( a^l [ W^l H_{v_{i},t}^l  \lVert W^l H_{v_{j},t}^l ]))}{ \sum_{k \in \mathcal{N}_{v_{i,t}} }{\rm exp}( \mathcal{F_\alpha} (a^l[ W^l H_{v_{i},t}^l  \lVert W^l H_{v_{k},t}^l]))}\\
H_{v_{i},t}^{l+1}&={\rm {MLP}} ( H_{v_{i},t}^l \lVert \sigma\sum_{j\in\mathcal{N}_{v_{i},t}} \alpha_{v_{i,j},t}^{l} H_{v_{i},t}^l )
\end{aligned}
\end{equation}
where $W^l$ is the weight matrix of a linear transition and $\lVert$ is the concatenation operation, $a^l$ represents a learnable vector, $\mathcal{N}_{v_{i},t}$ denotes neighbors of node $v_{i,t}$, $\alpha_{v_{i,j},t}^{l}$ is a learnable attentive coefficient of node $v_{j,t}$ to node $v_{i,t}$. $\sigma$ and $\mathcal{F_\alpha}$ are the sigmoid and LeakyReLU functions, respectively. 

We designed the above adaptive attention mechanism with the aim of identifying infected neighbors among all local connections, which are the most relevant nodes in local dynamic spreading. As noted above, the states of individuals might at the same time be determined by their own features (e.g., a recovered node at the current time step will remain in the same state and never be susceptible at the next time step). Therefore, we use a concatenation operation to keep the intact self-feature, rather than a weighted summation of self-feature and neighboring features like GAT. Finally, the last iteration layer $L$ outputs the node representation $H_t$ that distills the influence of local connections at each time step via the attention mechanism.

The microscopic governing rules $\mathcal{M}(t)$ can be effectively learned to illustrate the spreading process of the dynamic graphs based on the behavioral spreading in adjacent times:

\begin{equation}
\label{pred_trans}
\mathcal{M}(t)=\text{softmax}({\rm {MLP}}(H_{t} \lVert H_{t+1})).
\end{equation}

where ${\rm {MLP}}$ is a multilayer perceptron.

\subsection{Learn long-term dynamical features}\label{LDFTS}
\vspace{0.2cm}

Besides the governing rules for the contagion process through local network connections, effective learning of long-term dynamical features can iteratively refine features of the local governing rules, leading to their reinforced learning. Conversely, governing rules guide and refine the representation of neural networks. Therefore, we employ the GRU module to capture the short-term and long-term dependencies in the series of dynamic graphs.

The GRU~\cite{cho2014properties} model consists of two gates: the reset gate $R_t \in \mathbb{R}^{N\times h}$ and the update gate $Z_t \in \mathbb{R}^{N\times h}$, that are designed to explicitly control the information flow. The output $D_t$ of the GRU at time $t$, which combines the historical output $D_{t-1}$ and the current information $H_t^{'}$, can be written as:

\begin{equation}
\label{GRU}
\begin{aligned}
R_t&=\sigma (H_{t}^{'}W_{hr} + D_{t-1}W_{dr} +b_r),\\
Z_t&=\sigma (H_{t}^{'}W_{hz} + D_{t-1}W_{dz} +b_z),\\
\tilde{D}_t&= {\rm tanh} (H_{t}^{'}W_{hd}+(R_t\odot D_{t-1})W_{dd} +b_d),\\
D_t&=Z_t\odot D_{t-1} + (1-Z_t)\odot \tilde{D}_t.
\end{aligned}
\end{equation}

where $W_{h \cdot} \in \mathbb{R}^{d\times h}$ and $W_{d \cdot} \in \mathbb{R}^{h\times h}$ are the weight matrices, $b_{\cdot}$ is the bias vector, and $\odot$ denotes the Hadamard product operator. The reset gate determines the extent to which the essential knowledge of the previous time step should be retained, and the update gate determines the level of the previous hidden state $D_{t-1}$ brought into the next hidden state $D_t$.

The microscopic governing rules, $\mathcal{M}(t)$, of the SIR process contain mechanisms associated with both short-term (e.g., $\mathcal{N}_t^l$ ) and long-term (i.e., $\beta$) dependencies. In principle, $\mathcal{M}(t)$ can provide explicit guidance to the GRU module to facilitate the prediction of the next state. Thus, we incorporate the governing rules as an input to the GRU module:
\begin{equation}
\label{GRU_M_1}
\begin{aligned}
H_{t}^{'}&=[H_t \lVert \mathcal{M}(t-1) ], \\
D_t&={\rm GRU}(H_{t}^{'},D_{t-1}).
\end{aligned}
\end{equation}
The output $D_T$ of the GRU model at time step $T$ is further fed into an MLP layer to predict the future critical behavior $\hat{Y}$ over a period of time $T^{'}$:
\begin{equation}
\label{GRU_M_2}
\hat{Y}={\rm MLP} (D_T)  \in \mathbb{R}^{N\times T^{'} \times C}.
\end{equation}
where $C$ is the number of types of critical behaviors.

\subsection{Prediction for the system critical behaviors}
\label{cr}
\vspace{0.2cm}

To verify the effectiveness of our framework, we predict two representative quantities of the spreading dynamics in networked systems as shown in Figure \ref{fig:critical_behaviors_illustration}. In particular, we predict: {\it i)} the critical point (i.e., whether the system is above or below it, Fig. \ref{fig:critical_behaviors_illustration}a); and {\it ii)} the size of the outbreak (Fig. \ref{fig:critical_behaviors_illustration}b).

Our model can be optimized by the microscopic $\mathcal{M(t)}$ prediction loss and the $\hat{Y}$ prediction loss over $T$ dynamic graphs:
\begin{equation}
\mathcal{L} = l_m (\hat{Y},Y) + \lambda\sum_{t_0\leq t \leq T-1} l_b (\mathcal{M}(t),\mathcal{M}^{gt}(t)).
\end{equation}
where $Y$ and $\mathcal{M}^{gt}$ are the corresponding ground truth values, $l_m$ and $l_b$ denote a weighted cross-entropy loss that gives more weight to certain types of samples and Mean Square Error (MSE) loss, respectively. $\lambda$ is a hyper-parameter tuning the contribution of the governing rules.

\section{Results}
\subsection{Experimental Setup}\label{sec:setup}

\subsubsection{Datasets}

We evaluate the performance of our model on three different types of networked systems, namely, homogeneous, heterogeneous and real-world networks. For a given set $\mathcal{N}$ of nodes, Erd\H{o}s-r\'enyi networks~\cite{erdHos1960evolution} consider that any two nodes are connected independently with a probability $p$. The resulting network has a homogeneous (Poissonian) degree distribution. Scale free networks~\cite{barabasi1999emergence} represent those systems for which the connectivity pattern can be considered heterogeneous as far as the degree distribution is concerned. Finally, we also consider the real-world social network of Hamsterster \cite{nr}.

\subsubsection{Implementation details}

We generate a set of dynamic graphs as the model input according to the SIR spreading dynamics on networks of size $|\mathcal{V}| = 10^3$ with average degree $10$. The dynamic graph $G_{t_0}$ at the initial time $t_0$ has the state vector with one randomly chosen node to be infected and the remaining nodes are susceptible.

To train our model, we utilize a collection of dynamic graphs with $\beta$ ranging from $0.1$ to $0.9$ with a step of $0.1$ and $\gamma= 1$. For the real-world social network of Hamsterster, $\beta$ is taken from $0.02$ to $0.5$ with step $ 0.02 $. In our experiments, the dimension of the hidden layers of all neural networks is set to $32$. We set the batch size to $32$ and set epoch numbers to $100$. The model parameters that most accurately predicted the critical behavior of the validation set during the training are used to output the critical behavior $\hat{Y}$ of the test set. The Adam~\cite{Adam} method is employed for gradient descent optimization.

\subsection{Effectiveness in learning governing rules}

Before analyzing the effectiveness in predicting the system's critical properties, we evaluate the performance of our model in learning the governing rules $\mathcal{M}(t)$ of the spreading dynamics. We perform simulations on three different types of networks and compare the results with random and mean-field predictions. A random prediction considers that with probability $f$, a transition of one individual to a new state occurs, whereas it remains in the original state with probability $1-f$. Therefore, for the random prediction, the elements in $\mathcal{M}(t)$ are randomly chosen with such probabilities. For the case of the mean-field prediction, we set the elements in $\mathcal{M}(t)$ according to the mean probability of transition event across all dynamics graphs, effectively making a mean-field approximation for all individuals.

Table \ref{tab:accuracy_governing_rules} shows the prediction accuracy for the governing rules $\mathcal{M}(t), \ t_0 \leq t \leq T-1$ of the spreading dynamics learned from dynamic graphs. For all networks considered, our prediction accuracy is consistently better off by one to two orders of magnitude than that of the random (with $f=0.5$) and the mean-field predictions. The values shown in the Table indicate that, nonetheless the nonlinear functional form of the transition probability and the temporal and spatial properties of the set of infected neighbors $N^l_t$, the governing rules are effectively extracted from the adjacent dynamical graphs. Note that this is also a consequence of the proposed attention aggregation function, Equation~\ref{GAT}, that captures the most relevant interactions in the neighborhood of the infectious nodes.

\begin{table}[!htp]
\caption{Prediction accuracy for the governing rules $\mathcal{M}(t)$ of the spreading dynamics on three different types of networks.}
\begin{center}
\resizebox{0.95\linewidth}{!}{
\begin{tabular}{l|cc|cc|cc}
\toprule
\multirow{2}{*}{\diagbox[]{Models}{Datasets}} & \multicolumn{2}{c|}{Erd{\H{o}}s-R{\'e}nyi} & \multicolumn{2}{c|}{Barab{\'a}si–Albert} & \multicolumn{2}{c}{Hamsterster} \\ & MAE(\%) & MSE(\%)  & MAE(\%) & MSE(\%)& MAE(\%) & MSE(\%) \\
\midrule
Ours & 0.0675 & 0.0103 &  0.0556 & 0.0096 
& 0.0572 & 0.0111 \\
Random prediction & 0.4575 & 0.2201 & 0.4464 & 0.2136
& 0.4667 & 0.2271 \\
Mean-field prediction & 0.5306 & 0.3070 & 0.4079 & 0.2052
& 0.4151 & 0.2064 \\
\bottomrule
\end{tabular}
}
\end{center}
\label{tab:accuracy_governing_rules}
\end{table}

\subsection{Prediction accuracy for two critical behaviors}
\vspace{0.2cm}

\begin{figure}[!htp]\centering
\includegraphics[width=0.8\linewidth]{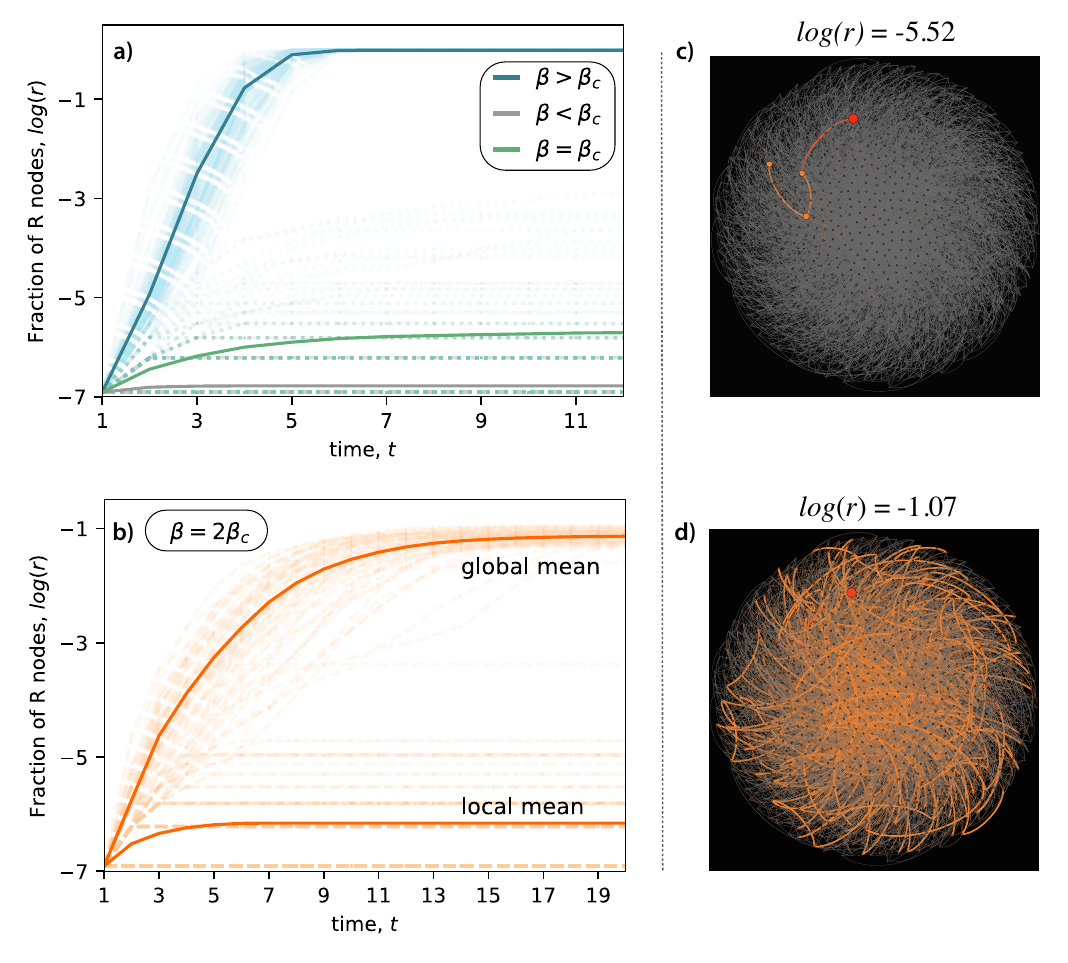}
\caption{Two representative properties characterizing the SIR spreading dynamics. Panel \textbf{a)} shows the evolution of the fraction, $r$, of recovered ($R$) nodes with time. Each solid line represents the average fraction of recovered nodes for spreading probabilities $\beta>\beta_c$, $\beta=\beta_c$, and $\beta<\beta_c$. The critical spreading point $\beta_c$ characterizes the transition point, above which the disease globally spreads, while below which there is no macroscopic outbreak. Panel \textbf{b)} shows the fraction of recovered nodes at $\beta=2\beta_c$, which corresponds to a regime of global spreading. For the same spreading parameters, the system can exhibit both microscopic and macroscopic outbreaks, as shown by the solid lines. Panels \textbf{c), d)} visualize the spreading outcomes on the network, corresponding to the steady state of two spreading trajectories in panel b). The SIR spreading dynamics is performed on a Barab{\'a}si–Albert network of $1000$ nodes with an average degree $10$. Each lighter dashed line shows one spreading realization (we performed a total of $1000$ realizations for each given set of parameters). Other parameters are $\beta=$ $0.01$, $0.05$, and $0.6$ in panel a) and $\beta=2\beta_c=0.05$ in b), respectively and $\gamma$ = 1.}
\label{fig:critical_behaviors_illustration}
\end{figure}

To evaluate the performance of our model, we inspect the local and global behavior of the system and use two metrics. The first is quantified by the number of recovered nodes ($R$ nodes) relative to the entire network size $|\mathcal{V}|$. The fraction, $r$, of $R$ nodes in the long-time limit depends on the spreading probability $\beta$ such that it is close to $0$ when the spreading is local (no macroscopic outbreak) while it approaches $1$ when $\beta$ grows, which indicates a state of global spreading (macroscopic outbreak). The critical point corresponds to the value at which a transition from one regime to the other takes place, see Figure \ref{fig:critical_behaviors_illustration}A. The second metric that we look at refers to the fraction of $R$ nodes that have a positive probability to be either in a local or in a global spreading regime, given that the system setting (the underlying network, the spreading parameters) is the same. Figure \ref{fig:critical_behaviors_illustration}B shows results for the latter metric. In what follows, we further evaluate the performance of our model in predicting the above two metrics that characterize the (critical) spreading dynamics.

\subsubsection{Ablation study}

To analyze the effectiveness of the learned governing rules, we first perform an ablation study of our model on three different networks. We first consider the GRU~\cite{cho2014properties} model, which makes evolution predictions solely based on a series of dynamic graphs without the guidance of governing rules. We then consider using only the aggregation function, ours-GNN, Equation~\ref{GAT}. The graph features $X=\{X_{t_0},\dots, X_{T}\}$ are sequentially input into Ours-GNN model and finally an MLP layer is employed to predict the critical behaviors. Table \ref{tab:ablation_study} shows the prediction accuracy of these model variations.
\begin{table}[!ht]
\caption{Ablation study of our model on three networks. Each column shows the classification accuracy (\%) of R-state nodes at the system's steady state.}
\begin{center}
\begin{tabular}{l|c|c|c}
\toprule
Models & Erd{\H{o}}s-R{\'e}nyi & Barab{\'a}si–Albert& Hamsterster \\
\midrule
GRU & 68.69 & 97.19 & 98.95 \\
Ours-GNN & 85.16 & 97.33 & 98.14 \\
Ours & 87.54& 98.03 & 99.49 \\
\bottomrule
\end{tabular}
\end{center}
\label{tab:ablation_study}
\end{table}

\subsubsection{ Local and global spreading}

\begin{figure}[!htp]\centering
\includegraphics[width=\linewidth]{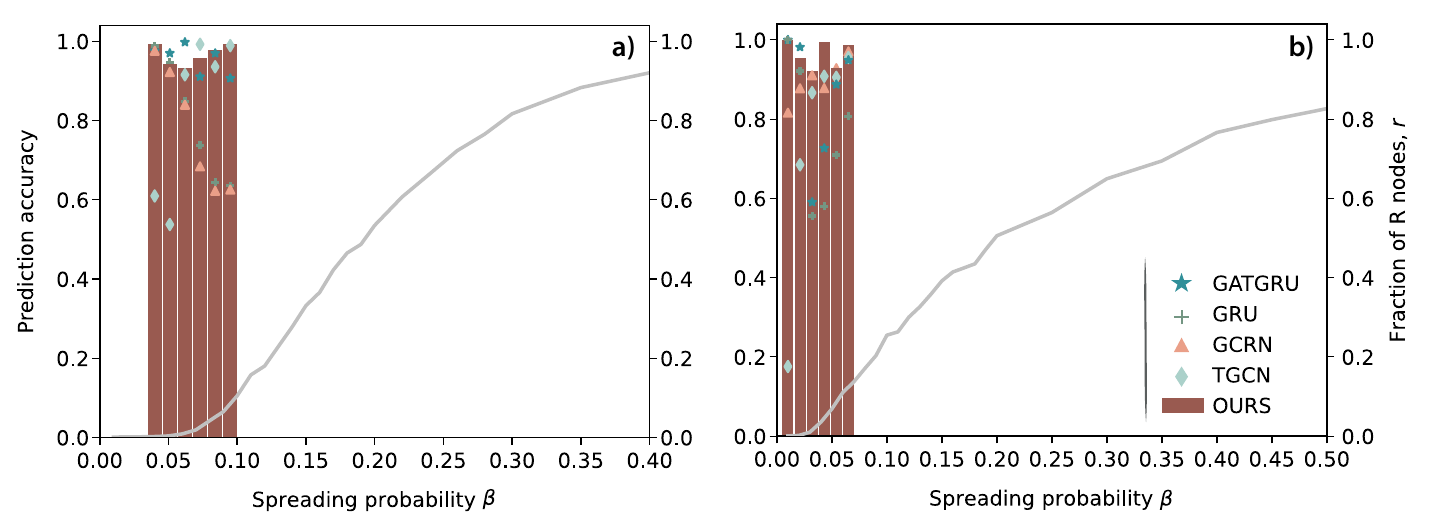}
\caption{Prediction of the critical transition from local to global spreading. Panels \textbf{A-B} show results on BA and the real-world social network Hamsterster. The gray curve shows changes in the fraction, $r$, of $R$ nodes as the spreading probability $\beta$ is increased. The critical point at which a transition occurs is $\beta_c$. The bar plots represent the prediction accuracy of our model for the fraction of $R$ nodes around the critical point. Markers show the prediction results of other models: GRU, GATGRU, TGCN, and GCRN. The spreading dynamics is performed on BA networks, which have an effective critical spreading probability $\beta_c = 0.05$.}
\label{fig:prediction_local_global_spreading}
\end{figure}

Figure \ref{fig:prediction_local_global_spreading} shows results for the first metric used to characterize the critical behavior of the spreading dynamics when it takes place on top of a BA heterogeneous network. We show the critical region around $\beta_c$ where a local spreading gradually transits to a global spreading, as shown by the curve of the fraction of $R$ nodes. Around the critical point, we plot the prediction accuracy $ 1 - \lVert \hat{r} - r\rVert/r$, obtained by the relative error between the target fraction $r$ of $R$ nodes and the predicted $\hat{r}$. Our model, shown as the bar plot, consistently exhibits a high accuracy in predicting the fraction of $R$ nodes near the critical point $\beta_c$. Admittedly, our results outperform GRU and GCRN across the examined values of the spreading probability on different networks. Though GATGRU and TGCN perform well on BA networks, they do not do so on the real-world social network Hamsterster.

\begin{figure}[!htp]\centering
\includegraphics[width=\linewidth]{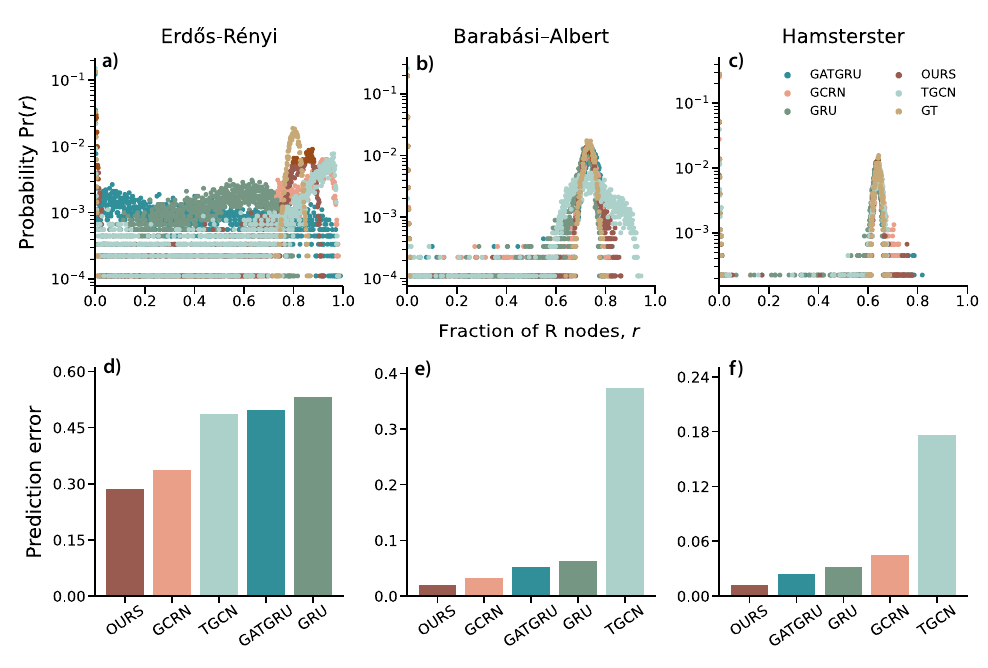}
\caption{Probability of having outbreaks of a given size. Panel (\textbf{A}) shows the predicted values for the probability of having a fraction $r$ of nodes in the state $R$ at the end of the spreading dynamics for fixed system settings. The ground truth (GT) consists of Monte Carlo simulations of the SIR dynamics. The underlying network is an ER with size $|\mathcal{V}|=1000$ and average degree $10$. Each of the realizations of the spreading dynamics is started using the same initial state $X_{t_0}$, using $\beta = 2\beta_c$. Panels (\textbf{B}-\textbf{C}) display the same prediction results, but on BA networks and the real-world social network used throughout this study, with $\beta = 4\beta_c$, $\beta = 8\beta_c$, respectively. Panels (\textbf{D}-\textbf{F}) show the Jensen-Shannon divergence for the predicted distribution and the ground truth distribution, for the cases displayed in panels A to C, respectively.}   
\label{fig:prediction_two_peaks}
\end{figure}

\subsubsection{Outbreak size distribution}

The transition that takes place at $\beta_c$ is of second order, which means that it is a smooth transition. However, due to the stochastic nature of the spreading dynamics, and the fact that the SIR model has infinitely many absorbing states (in the thermodynamic limit), the final size of an outbreak is hard to predict, even if it is a key magnitude from an epidemiological point of view. In other words, for the same set of spreading parameters and network structures, the system's dynamics can either evolve into a global spreading macrostate (where the fraction of $R$ nodes is large) or die out after a local spreading (i.e., $r$ is close to $0$), see the example shown in Figure \ref{fig:critical_behaviors_illustration}b) for $\beta = 2\beta_c$. This makes accurate predictions hard.

Figure \ref{fig:prediction_two_peaks} shows the performance of our model in predicting the probability of having an outbreak of a given size (e.g., of having a fraction, $r$, of nodes in the state $R$ at the end of the spreading process) for the three types of networks discussed before and different values of $\beta$. Top panels (Fig. \ref{fig:prediction_two_peaks}A-C) show results for the previous probability of having an outbreak of size $r$. For \cmt{the} sake of completeness, we also show results for the other models we compare with: GATGRU, TGCN, GCRN, and GRU. Finally, the bottom panels of Figure \ref{fig:prediction_two_peaks}D-F display the computed prediction error using Jensen–Shannon divergence to quantify the similarity between the targeted distribution and our prediction. The smaller the value of the Jensen–Shannon divergence the higher the similarity between our predicted distribution and the ground truth distribution (i.e., the distribution from Monte-Carlo simulation of SIR dynamics). As it can be seen, our model outperforms other models and prediction accuracy increases as one moves away from the critical point (which is where bigger fluctuations happen).

\section{Conclusions.}

In this paper, we have proposed a new graph neural network-based framework to characterize and predict key properties of contagion dynamics (as given by the SIR model) in networked systems. To overcome the difficulties in predicting the critical properties that typically characterizes the system's transition between different states in a narrow region of parameters, we propose a graph neural network-based module to first learn the rules governing the system evolution. Leveraging this learned knowledge, we refine and guide the neural network learning long-term dependencies from a series of dynamic graphs, the combination of which facilitates the prediction of critical properties. We have shown the effectiveness of our method using two key epidemiological observables through extensive numerical experiments on various types of networks. Comparisons of our methodology with other methods available in the recent literature show higher accuracy and better performance of the proposal presented here. Altogether, our study demonstrates via the scenario of contagion dynamics that effective learning of the underpinning hidden physical mechanism empowers the representability of graph neural networks, which may steer further applications such as predicting pattern formation and system instabilities.

\section*{Acknowledgments}
X. W. was supported in part by GuangDong Basic and Applied Basic Research Foundation (Grant No. 2025A1515011389, 2023B0303000009), and in part by National Natural Science Foundation of China under Grant 62327808. Y. M. was partially supported by the Government of Aragon, Spain, and ERDF "A way of making Europe" through grant E36-23R (FENOL), and by Ministerio de Ciencia e Innovación, Agencia Española de Investigación (MCIN/AEI/ 10.13039/501100011033) Grant No. PID2023-149409NB-I00. The funders had no role in the study design, data collection, analysis, decision to publish, or preparation of the manuscript.


\end{document}